Title: Droplet size and velocity measurements in a cryogenic jet flame of a rocket-type combustor using high speed imaging.

Authors : Nicolas Fdida, Lucien Vingert, Arnaud Ristori and Yves Le Sant.

Affiliation : Onera – The French Aerospace Lab

Corresponding author : nicolas.fdida@onera.fr

Affiliation : Onera, the French Aerospace Lab, FR-91123, Palaiseau, France

Abstract :

Injectors of cryogenic liquid rocket engines produce large polydisperse and dense sprays due to the pressure and mass flow conditions. Atomization is the dominating process that drives the flame behaviour in cryogenic jet flames, when the propellants are injected in subcritical conditions. The main objective of this study is the characterization of a liquid oxygen (LOX) spray in gaseous hydrogen (GH2), in reacting conditions. In the breakup region where liquid particles are not spherical, laser based drop size techniques suffer from a low validation data rate, thus imaging techniques can be better suited to characterize the spray. High speed shadowgraphs were used to provide the spray characteristics such as sizes and velocities of the LOX dispersed phase atomized by a GH2 co-flow injected by a shear coaxial injector in a 1MPa combustion chamber. Cryogenic combustion investigations presented in this paper were carried out on the Mascotte test bench, at Onera. The reacting case was compared qualitatively to a cold flow test, with gaseous He instead of GH2, for which LOX spray shadowgraphs were also recorded. Velocity of the dispersed phase in reacting conditions was obtained with two different imaging methods, which were applied to the same shadowgraphs: a PTV algorithm and a PIV software, developed at Onera. Droplet sizes were measured by image processing in the atomization zone and results were compared with Sauter mean diameters obtained with phase Doppler measurements from the literature. The droplet size is combined with PTV to obtain droplet size/velocity correlations which show the influence of the flame on the droplet size evolution.

1. Introduction

      Detailed experimental studies of cryogenic propellant combustion, in well controlled and nevertheless representative operating conditions, are needed to optimize the design of high performance liquid rocket engines such as the Vulcain 2 of the Ariane 5 launcher and the VINCI engine which will propel the future Ariane 5 ME and Ariane 6. Several research facilities, such as Mascotte (Onera), Pennstate University test bed or the P8 (DLR) were built to provide hot fire tests, under realistic injection conditions. In this context, the research facility called Mascotte was built up by Onera twenty years ago, to investigate elementary processes that are involved in the combustion of liquid oxygen (LOX) and gaseous hydrogen (GH2) near the exit of a single coaxial injector (see Vingert et al., 2000). In a standard cryogenic flame, propellants are injected by a shear coaxial injector composed of a central low-velocity LOX jet surrounded by an outer high velocity co-flowing GH2. The flame structure has been studied on the Mascotte test bench by Habiballah et al., 2006 and Candel et al. 2006 in various operating conditions. Two different regimes of the flame can be considered, depending on the combustion chamber pressure $Pc$ with respect to the critical pressure of oxygen $p_{c(O2)}$. In supercritical conditions, LOX is injected in a supercritical state where a turbulent mixing process is prevailing in the flame behaviour. At a lower chamber pressure, smaller than $p_{c(O2)}$, combustion is considered as subcritical. In this case, atomization is one of the most important processes that drive the flame. In this regime, LOX forms a dense liquid jet that breaks into ligaments, which are secondly atomized by the aerodynamic shear forces of the high velocity outer gas flow. There is a need of experimental data in these operating conditions to improve the physical knowledge and modelling for liquid rocket combustion, as mentioned by Benjamin et al. 2010. Several authors (Locke et al., 2010, Davis and Chehroudi, 2007, or Oschwald et al., 2006) have reported experimental data in subcritical and supercritical conditions, with a shear coaxial injector to obtain the LOX jet penetration length in various operating conditions. Those studies have lead to correlation laws predicting the LOX penetration length as a function of physical parameters such as the momentum flux ratio $J$, or dimensionless Reynolds or Weber numbers. Davis and Chehroudi, 2007, proposed two different correlations, depending on whether the spray is in subcritical or supercritical conditions. In the subcritical regime, the atomization process can be divided in two steps: the primary atomization of the dense LOX jet is driven by large amplitude oscillations producing large liquid elements (Matas and Cartellier, 2010). From the injector exit, the liquid jet is also peeled-off at its surface creating ligaments of scale smaller than the liquid jet diameter (Marmottant and Villermaux, 2004). The secondary atomization is the process that breaks those liquid fragments into droplets or ligaments, as illustrated by Mayer and Tamura, 1996. We propose here to study the secondary atomization zone of the spray, which gives birth to a wide range of droplet sizes, which are subsequently vaporized by the flame. There is a lack of data on drop size and velocities of a cryogenic jet flame. These data are obtained relatively close to the injector ($6<x/d<12$), and off-axis ($2<r/d<4$), which complete the existing datase.

      Phase Doppler Particle Analyzers (PDPA) have been used in dense sprays since the mid eighties (Kraemer and Bachalo, 1986) and are now commercially available to measure the droplet size and velocity in dense sprays. Gicquel and Vingert, 2000, have investigated the spray studied here, on the Mascotte test bench but the atomization zone was not deeply investigated because of a low validation rate of the size measurements. Indeed, in this region where lots of liquid elements are not spherical, Phase Doppler systems have difficulties to measure large droplet sizes with confidence. Damaschke et al., 1998, showed that the droplet morphology is a critical parameter to obtain a reliable drop size. Moreover, Rousset et al., 2009, pointed out that the optical setup of a Phase Doppler system needs some cautions to be applied to cryogenic two-phase flows, particularly due to the relative refractive index of the

medium. Imaging techniques are suited to perform drop-size measurements in the atomization area of a dense spray because they are not affected by the non spherical shape of the liquid structures (Fdida and Blaisot, 2010). Moreover imaging diagnostics can be less sensitive to index variations due to the high temperature gradient between the flame and the cryogenic fluid. The use of high-speed cameras can also provide information on the jet flame dynamics: the penetration length (Fdida et al. 2013), the combustion instabilities (Mery et al. 2013, Hardi et al., 2014) or the droplet sizes and velocities as in this present study. As the goal is to characterize the secondary atomization, we propose here to study the dispersed phase velocity with two different methods applied directly to the shadowgraphs of the dense spray. The first one is a Particle Tracking Velocimetry (PTV) method developed by Malek, 2008. The second method is a PIV algorithm named FOLKI-SPIV, developed at Onera by Champagnat et al., 2011, which was directly applied to the spray shadowgraphs. Goldsworthy et al., 2011, have shown that PIV based algorithms can lead to reliable measurements of the dispersed phase. In this case, droplets are natural tracers of the dispersed phase. Those methods are based on different kinds of algorithms, providing complementary information, each method having its own advantages and limitations, which will be addressed in the paper.

The experimental setup is first presented to describe the Mascotte test bench, the optical setup and the image processing methods. Experimental results presented in this paper were obtained in the stationary phase of the reacting flow. A qualitative comparison is performed between the reacting case and the cold flow test. The mean drop size diameters and the probability density functions were analyzed in several zones of the spray by image processing. Droplet size/velocity correlations were obtained with PTV. The velocity of the dispersed phase was obtained with PTV and compared with results of FOLKI-SPIV.

2. Experimental setup

    2.1 Mascotte test bench

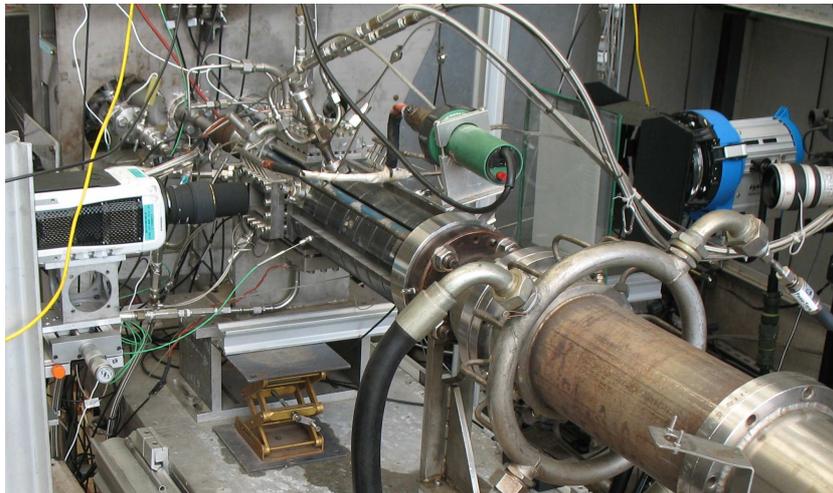

Figure 1: Overview of the Mascotte test bench equipped with imaging setup.

The Mascotte test bench, located at Onera Palaiseau center, was designed for experimental studies of cryogenic combustion. It provides data on elementary processes occurring in cryogenic jet flames such as atomization, vaporization and turbulent combustion, in conditions representative of actual rocket combustors. Several versions of the hardware have been developed and manufactured for the different items of interest in experimental research (Habiballah et al., 2006 and Candel et al., 2006). In the version used for the study reported here, a visualization module with two optical ports was placed at the injector exit. The combustion chamber had a square cross section of 50 mm side and was about 500 mm in length. A transition segment precedes the nozzle, to cross-connect the square combustion chamber to the axisymmetric nozzle. Ignition is provided by a gaseous oxygen/ gaseous hydrogen torch igniter placed in the first ferrule. An overall view of the Mascotte test bench is shown on figure 1, the camera is located on the left side, the light source on the right, illuminating the combustion chamber perpendicularly to the injection axis.

The combustion chamber was fed with gaseous hydrogen and liquid oxygen through a single coaxial injector. The injector was constituted of a central tube of diameter $d$ and a coaxial outer tube fed with hydrogen of outer diameter $2.4d$ and inner diameter $1.12d$. In order to prevent the heating of the LOX during injection, the injection head was cooled by liquid nitrogen. Thus LOX was injected at ~90 K while GH2 was injected at room temperature. To protect the windows from hard thermal shocks, a film cooling of helium was injected with a mass flow rate of ≈6g/s. Experiments were performed on the A-10 operating point, defined by the research group on "combustion in rocket engines", referenced in Candel et al., 2006 and Habiballah et al., 2006. The chamber pressure was ≈1.0 MPa, the mixing ratio $ROF$ was ≈2 and the momentum flux ratio $J$ was ≈13 . Considering the oxygen critical pressure $p_c(O2)$=5.04 MPa and temperature $T_c(O2)$=154.6K, LOX was injected in the combustion chamber in subcritical conditions.

|  | Fire test | Cold flow |
|---|---|---|
| $Pc$ (MPa) | 1.00 | 0.96 |
| $Tinj(LOX)$ (K) | 89 | 90 |
| $Tinj(GH2)$ (K) | 296 | 290 |
| $Re_d$ | 62300 | 64300 |
| $ROF$ | 2.4 | 1.7 |
| $J$ | 12.99 | 12.38 |
| $We_G$ | 26400 | 25300 |

Table 1: Main operating parameters during the steady-state conditions.

Table 1 summarizes the main parameters of cold flow and hot fire operating conditions. Those average values were computed over the whole set of test runs for this study. Comparison between the cold tests (H$_2$ line fed with He) and firing tests (H$_2$ line fed with H$_2$) seemed satisfactory. The similarity between cold and reacting case was based on the following assumptions: the geometry of the chamber, the momentum flux ratio $J$, the chamber pressure $Pc$ and the LOX mass flow rate were conserved. By replacing H$_2$ by He, the molar weight is doubled and assuming that He and H$_2$ are following the perfect gas law and that both temperature and pressure are kept constant, the density is also doubled. To keep $J$ constant, the injection velocity of He had to be decreased compared to the reacting case with H$_2$, to counterbalance the density increase. Therefore the helium injection velocity was set at $1/\sqrt{2}$ times the H$_2$ injection velocity and thus the He mass flow rate was equal to the H$_2$ mass flow rate times $\sqrt{2}$.

The gaseous Weber number $We_G$ was ~25000 and ~26000 for the cold and reacting cases respectively. The liquid Reynolds number $Re_d$ based on the LOX post diameter $d$ was ~64000 and 62000 for the cold and reacting cases respectively. Lasheras and Hopfinger, 2000, proposed a classification of coaxial jet breakup regimes as a function of three parameters, $We_G$, $Re_d$ and the momentum flux ratio $J$. In our study, the jet atomization could be considered as a fibre-type breakup. In this regime, the atomization is driven by the gas velocity coming from the outer ring of the coaxial injector. The relative motion with respect to the atomizing gas causes an instability of the liquid stream and a tendency to form waves that grow rapidly in amplitude to the point that the liquid stream breaks down into large liquid structures. Droplets and ligaments are created by the relative motion between the liquid and the atomizing gas (Vingert et al., 1995).

The time origin *t0* of the acquisition sequence was defined as the beginning of the recording sequence of the physical sensors on the test bench (temperature, pressure, flow meters…). Typical pressure and flowrate traces are shown on figure 2, each cold flow test lasted 60s whereas the hot fire test was only 50s, with a stationary period of the flow of about 10s and 20s for hot and cold tests, respectively. Vertical dashed lines show the beginning of the three recording periods of the camera of ≈0.5s each, inside the stationary time period of the flow. *Pc* was nearly the same and kept constant for both cases but instabilities were often noticed on the *Pc* signal for the reacting tests only, between [t0+34; t0+39]s. Thus this time period was not taken into account. 14 hot runs and 21 cold flow tests were conducted to obtain a complete map of the spray. In the following, all presented measurements were made during the stationnary period of the flow, with the last recording period, which was [t0+47; t0+47.5]s and [t0+43; t0+43.5]s for cold and hot runs respectively.

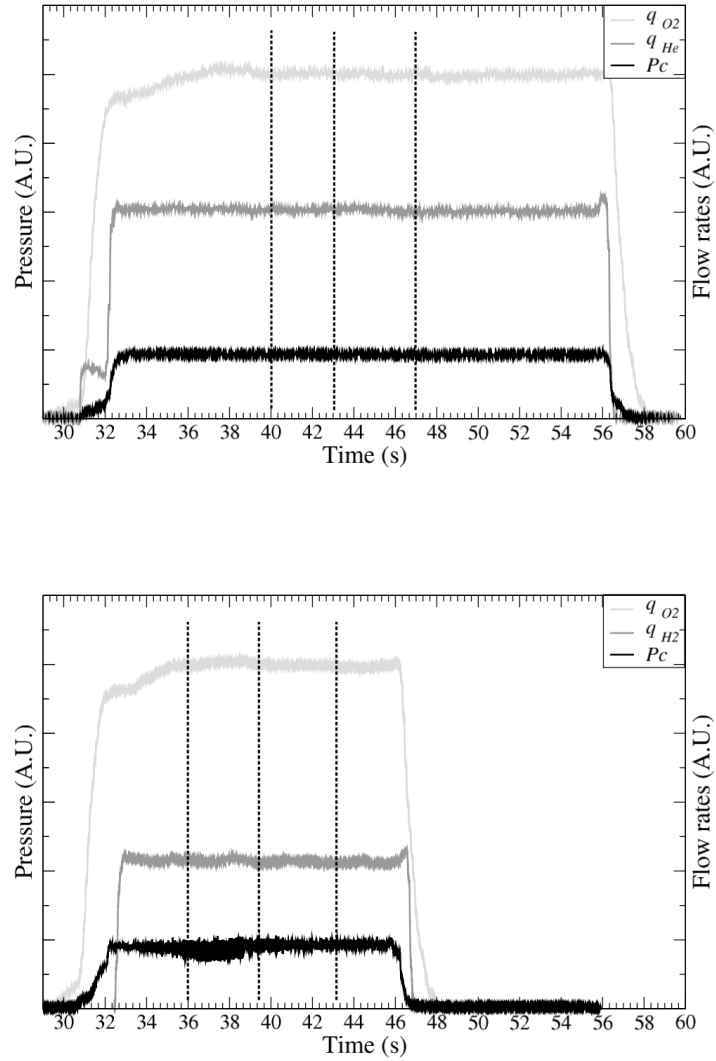

Figure 2: Typical pressure and flow rates signals for a cold flow (top) and hot fire test (bottom). Vertical dash lines indicate the beginning of the recording periods

2.2 Shadow imaging setup

The spray was enlightened in a backlight configuration by a high magnitude light source Prolight 575W, providing a white incoherent and continuous light. A Fresnel lens and a transparent sheet were placed in front of the light to ensure a homogeneous background on images. The dark level of the camera was subtracted for each run. Images of the spray were recorded with a Phantom v711 high-speed camera. This camera is composed of a 12-bit CMOS sensor of 1280 x 800 square pixels of 20 µm side. The size of the sensor is directly related to the frequency rate. We used three partitions of the memory, each one with its own frame rate/sensor size setup, starting at the times shown by vertical dashed lines on figure 2. The resolution was varying from 1280 x 800 pixels to 512 x 384 pixels, depending on the frame rate. For the reacting case, the highest frame rate was 25 kHz, with a 512 x 512 window, allowing to track droplets in two successive images. For the cold tests, a higher frame rate of 33 kHz was used with a 512 x 384 window but this frequency was not rapid enough to follow droplets. An objective of 105 mm focal length was mounted on the camera and the aperture was fixed to f/5.6 during the whole tests. The exposure time of the camera was fixed at its minimum to 1µs to freeze the droplets on images. It can induce a pixel shift on droplet images, due to the droplet velocities, which was estimated at ¼ of a pixel in the reacting conditions. The optical setup was kept the same for cold and hot fire tests.

To cover the whole window, whose length is ~15*d*, the camera was moved in a vertical focus plane (x,y) with small fields of view of 2*d* x 2*d*. The *x* axis is considered as the injection axis. The small size images were assembled on figure 3, the more rigorously possible, to compare the reacting and the cold jet at the same scale. We can see that the result is quite homogeneous, except on the border of the window where light conditions were different from those in the center, due to the presence of the window flanges, cutting a part of the light. We used those borders to have a similar view for both conditions (cold and hot). For the reacting case only, we define four 512 x 512 pixels regions of interest (ROI): A, B, C, D, shown on figure 3, which were used to perform drop-sizing and velocity measurements. Those ROI have been chosen off-axis, focused on the secondary atomisation zone, where the optical density of the spray was not too high so that we could characterize droplets individually and perform image processing methods.

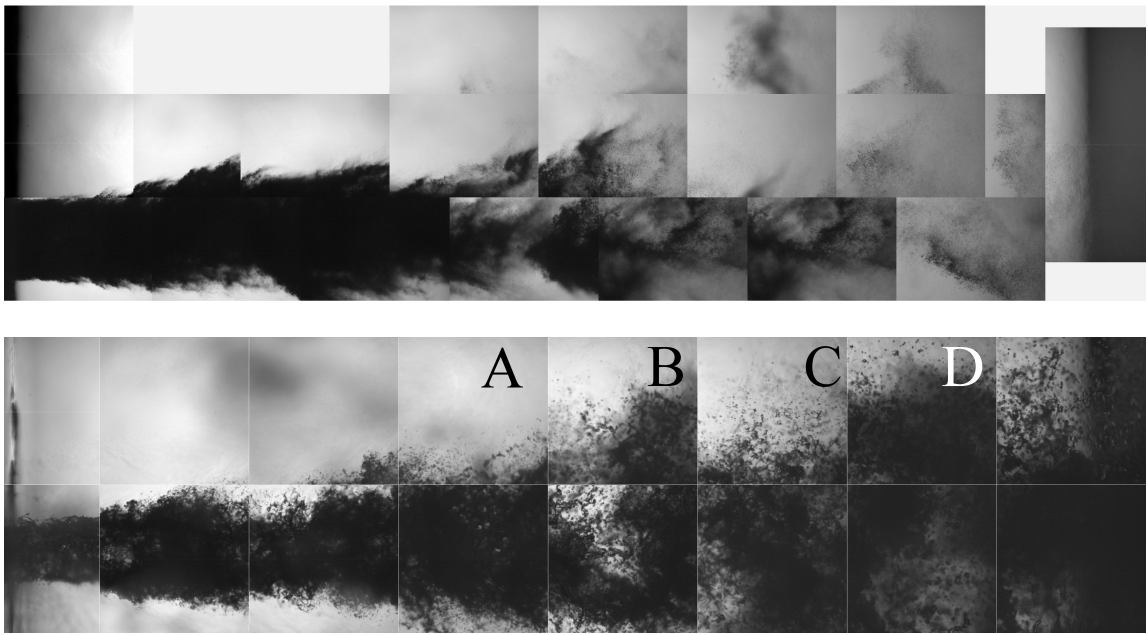

Figure 3: Instantaneous images of the spray in cold flow (top) and reacting conditions (bottom). Field of view: 15*d* x 4*d*.

2.3 Drop-sizing imaging method

The drop-size of the spray was obtained by an image processing algorithm developed by Fdida and Blaisot (more details in Fdida and Blaisot, 2010, and Fdida et al. 2010) and Blaisot and Yon, 2005, consisting in three steps. First, images were normalized to enhance the contrast and to correct non uniform background illumination. Then images were binarized by a double threshold technique: a classical threshold based on the gray level range in the image and a threshold based on wavelet transform (in order to detect out-of-focus droplets). This technique detects local gray level variations of low-contrast images in order to enhance the number of detected droplets. Each droplet is then associated to a surrounding mask, separated from its neighbours in order to be analyzed individually. The subpixel contour of the droplet is then computed and its diameter is calculated as the equivalent surface diameter of the binarized droplet image at a level equal to 61% of the local gray level amplitude, according to the imaging model. The minimum measurable diameter $D_{min}$ = 43.3 µm, corresponding to a few pixels, was obtained by the calibration procedure described below. $D_{min}$ appeared to be small enough to measure the smallest droplet sizes in the hot spray, otherwise the probability density functions (*pdf*), presented further in section 3.3, would be truncated towards the small diameters. The drop-size algorithm also provides the coordinates of the gravity center of the projected area of the droplet image, which are also used as an input of the PTV algorithm.

The use of a high speed recording rate implies that the same droplet can be measured on several successive images. As one of the objectives of this work was to obtain an independent droplet sizing (but also the droplet velocities), we didn't perform the drop-size algorithm on all images to avoid measuring the same droplet several times. Thus we choose to run the drop-sizing algorithm every 25 images, that is every ms, at a frame rate of 25 kHz. We estimated that a droplet can travel half of the image during this period. Thus, for each test run, a total number of 577 images of 512 x 512 pixels were processed from ROI A to D. The number of droplets counted for each measurement location was at least 7000 droplets. The average number of measured droplet per image varied from 12 to 17, depending on the location of the measurement from ROI A to D respectively.

The main advantage of the method is to provide reliable diameter estimation over a wide range of out-of-focus droplets, which allows counting more droplets than most of drop-size imaging methods. In such experiments for which the effective stationary running time is short, this method is useful to provide a statistically robust amount of data, with a minimum of images. Nevertheless, to obtain such a diameter correction for out-of-focus droplets, it is necessary to perform a calibration procedure. This drop-sizing algorithm was already used on dense sprays produced by diesel and gasoline automobile injectors, with good comparisons with others optical drop-sizing methods (see Blaisot and Yon, 2008 and Fdida et al., 2010).

A first calibration curve was necessary to estimate correctly the diameter of the drops, independently from their depth of focus (Fdida and Blaisot, 2010). The calibration procedure consists in recording images of calibrated discs engraved on a glass substrate for several distances from the focus plane of the optical system. The ratio of the real diameter a to the measured value $r_{meas}$ is directly related to the contrast $C_0$ of the droplets, as shown in figure 4. By comparing the dotted line and the solid line, we can say that experimental data were in a good agreement with the model. This regression curve is then used by the image processing algorithm to estimate the real diameter of droplet images.

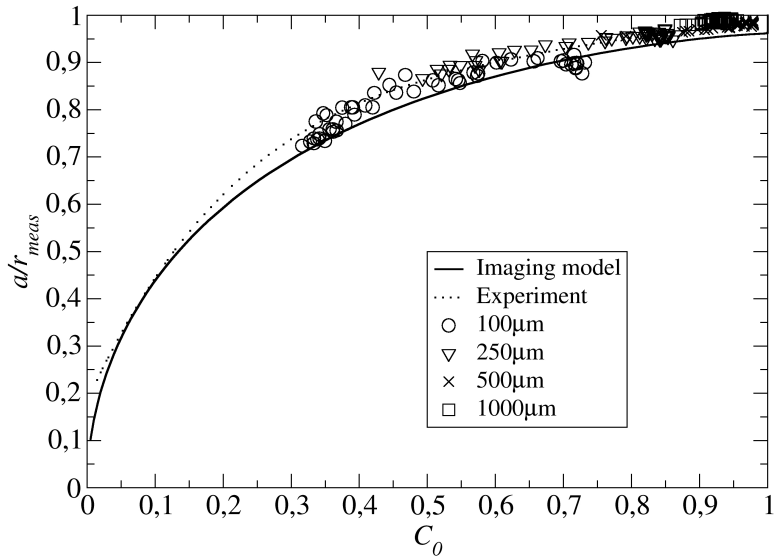

Figure 4: Calibration of the sizing procedure with calibrated discs. $a/r_{meas}$ is the ratio of the actual diameter $a$ to the measured value $r_{meas}$, represented here as a function of the contrast $C_0$. The solid line indicates the imaging model calculation and the dotted line indicates the regression curve from experimental data.

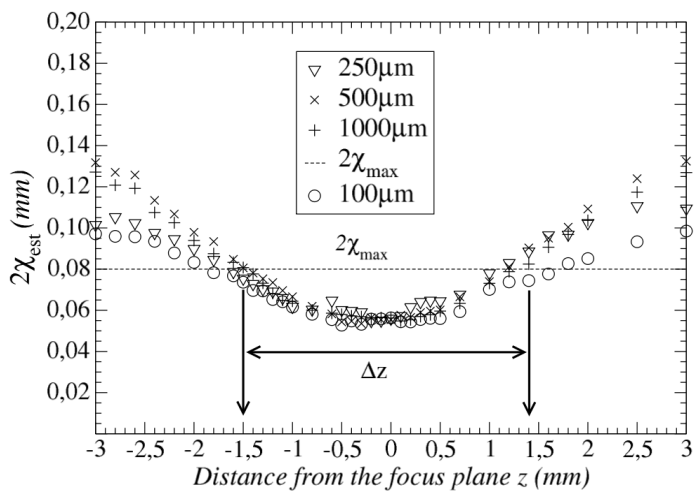

Figure 5: Calibration of the *psf* half width $\chi_{est}$ along the optical axis $z$ with calibrated discs. $\Delta z$ indicates the depth of field of the imaging system. $\chi_{max}$ is the depth-of-focus criteria.

In order to sort the droplets relatively to their spatial position from the focus plane, the level of out-of-focus of each droplet was obtained thanks to a second calibration curve, presented on figure 5, which characterizes the point spread function (*psf*) of the imaging system. Indeed, the *psf* is related to the position of droplets of any size relatively to the focus plane. The *psf* half width $\chi_{est}$ was estimated for each droplet from relations derived from the imaging model, proposed by Fdida and Blaisot, 2010. The depth-of field $\Delta z$ of the drop-sizing method was obtained by choosing the maximum *psf* width $2\chi_{max}$ =0.08 mm, indicated by the horizontal dashed line on figure 5. The criteria for $\chi_{max}$ was chosen on a range where the *psf* half width $\chi_{est}$ of the calibrated discs varied in the same way, that is assumed in the out-of-focus range -1.5 mm<$z$<1.5 mm. Otherwise a higher $\chi_{max}$ could lead to an important bias for the smallest droplets in the spray. Droplets of low contrast ($C_0$ <0.1) and high level of out-of-focus ($\chi_{est}$ <$\chi_{max}$) were so rejected. For practical reasons, the calibration experiments were performed outside of the test bench, with the same optical setup, with light passing through the window of the test chamber before enlightening the calibrated discs.

2.4 Velocity techniques: PTV and FOLKI-SPIV

2.4.1 PTV technique

The PTV system provides the velocity of individual droplets detected by the drop-sizing method previously described, knowing the time interval $\Delta t$ between two images. The PTV algorithm was originally developed by Malek, 2004, for characterizing 3D flows recorded by holography and was adapted by Fdida, 2008, to 2D flows. 2D subpixel coordinates of the droplets on a couple of successive images, obtained with the drop-size algorithm, are the input of the algorithm. The algorithm finds the best matching between the two sets of points. The output of the algorithm is the displacement of each matched particle in a pair of images.

For the same reason of independency on the droplet samples, only one image pair every 25 images was processed to avoid measuring the same droplet several times. Thus the same 577 images were processed by the PTV and the drop-size algorithm, except that PTV needs two successive images instead of one for the drop-sizing method. An example of the vectors found in an instantaneous image pair with the PTV algorithm is shown in figure 6.a.

The time interval $\Delta t$ is a relevant parameter to ensure the best possible matching. The following criterion must be verified to choose the optimum $\Delta t$ for a given flow: 0.4<$\Delta d_t/\Delta d_m$<0.8. Where $\Delta d_t$ is the average distance covered by a particle between an image couple and $\Delta d_m$ is the mean distance between two particles in a single image. In other words, if the average displacement is greater than the average distance between droplets, the PTV algorithm has difficulties to find the actual matching. $\Delta d_m$ was deduced from the drop-sizing results, by the relation : $\Delta d_m = \sqrt{(2S/(N_{d/im}\sqrt{3}))}$, where $S$ is the image surface and $N_{d/im}$ is the droplet number per image. $\Delta d_t$ was obtained by multiplying $\Delta t$ with the estimation of the mean droplet velocity. We found that the criterion was valid for $\Delta t = 0.08$ ms and $\Delta t$=0.04 ms. The PTV algorithm efficiency is evaluated by the percentage of matching droplets over the total number of droplets in a pair of images. The best matching percentage was obtained with the smallest recording time interval $\Delta t$=0.04ms which corresponds to the maximum recording frequency. With this setup, the average matching percentage was about 60% of the input particles. This percentage was too low to obtain a minimum of 5000 droplet displacements in each of the 577 images pairs obtained in ROI A, B, C and D, with the $\chi_{max}$ criteria for drop sizing given in the previous section. To obtain a larger sample of droplets, the out-of-focus criteria $\chi_{max}$ was adapted to detect more droplets, and thus $\chi_{max}$ was set at 0.15 mm. Moreover, about 10% of the image pairs processed by the PTV algorithm resulted in a wrong displacement. It means that the average displacement within the image pair was in the wrong principal flow direction ($U<0$), thus in this case the entire vector field was rejected. Moreover we eliminated all image pairs, which presented a low matching rate, i.e. when less than 50% of droplets were matched.

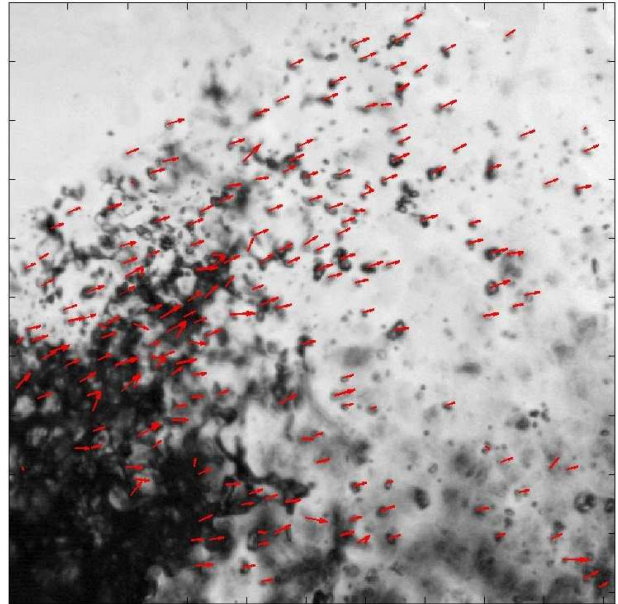

a)

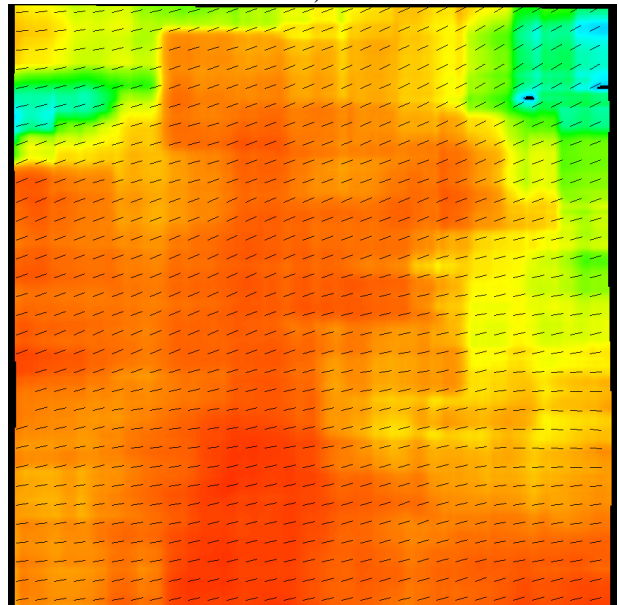

b)

Figure 6: Illustration of the PTV and FOLKI-SPIV methods on the same image pair, $\Delta t$=0.04 ms, Field of view: 2$d$ x 2$d$. *a)* Normalized image, ROI B superimposed with the vectors (red) found by PTV. b) FOLKI-SPIV vector field (black lines) superimposed on the s*core* colormap.

### 2.4.2 FOLKI-SPIV

FOLKI-SPIV is developed at Onera by Champagnat et al., 2011. It is an extension of FOLKI, Le Besnerais and Champagnat, 2005, which is an optical flow software, based on the Lucas-Kanade paradigm, 1981. FOLKI-SPIV is found comparable to that of state-of-the-art FFT-based commercial PIV software, while being 50 times faster. Indeed, the computation of dense PIV vector fields uses the Graphics Processing Units (GPUs) resources of the computer. This method is based on an iterative gradient-based cross correlation optimization method, which provides an accurate and efficient alternative to multi-pass processing with FFT-based cross correlation. Density is meant here from the sampling point of view (one vector per pixel is obtained), since the presented algorithm, FOLKI, naturally performs fast correlation optimization over interrogation windows with maximal overlap. The processing of the same 577 image pairs was achieved in less than 15 minutes on a NVIDIA GeForce GTX 480, while the PTV algorithm spent a couple of hours for the same job.

This algorithm is usually used for PIV images to measure velocity fields of seeding particles in a flow (thus white spots on a black background). Here droplets are used as natural tracers of the liquid flow, marked as black spots on backlighting images (see figure 6.a). The algorithm applied on such backlighting images principally detects the droplet movements in areas where the liquid/gas interface is distinct. On the spray edges, like in ROI A and B, where few droplets were detected, FOLKI-SPIV still captures movements of patterns due to either rapid variation of refractive index in the gas phase or of the few droplets remaining. Those patterns, barely noticeable on normalized images (but not on figure 6.a), are enhanced by the FOLKI-SPIV pre-processing. This area is composed of vaporizing oxygen gas and GH2, mixed by the turbulent flame, causing light deviation, which can be seen on images, like in a Schlieren optical setup.

The algorithm confidence is evaluated with the normalized cross correlation *score* as used in standard PIV software. It is ranging from 0 (no correlation at all) to 1 (perfect correlation). It was found particularly low in areas suffering from a lack of drops (black and blue regions on figure 6.b). Areas exhibiting low correlation values are not reliable, thus pixel showing a correlation value below 20% were not taken into account (transition from blue to green on colormap of figure 6.b). FOLKI-SPIV is adapted to enable the processing of PIV images, with the possibility to handle borders and masks.

FOLKI-SPIV provides one velocity vector per pixel, but the effective spatial resolution remains linked to the interrogation window size, as for conventional PIV methods. The optimal interrogation window size in term of cross correlation was found to be 99 x99 pixels, which was large regarding the image size (512 x 512). As images were not homogeneously filled with droplets, a large interrogation window size enabled to obtain more reliable results in areas suffering from a lack of particles. Increasing the interrogation window size provides larger spatial scales but it also decreases the signal to noise ratio.

# 3 Results

## 3.1 Cold/hot flow comparison

The morphology of the jet at the injector exit is shown in figure 7: the cold flow (LOX with gaseous He, figure 7.a and 7.b) and the reacting flow (LOX with gaseous $H_2$, figure 7.c and 7.d) can be qualitatively compared. For each case, a couple of successive images (1280 x 800 pixels) are shown, recorded at 7.5 kHz with the same optical setup. The LOX post exit of the injector is located on the left edge of the images.

The Reynolds and Weber numbers shown on Table 1 are similar in both cases and indicates that the LOX spray is in a fibre-type break-up regime, which is characterized by the creation of very thin and short liquid fibres created as soon as the continuous liquid jet exits from the nozzle. These fibres are rapidly peeled off the jet, stretched by the differential velocity between the liquid jet and the outer gas stream. Due to this stretching, the liquid jet is highly modified and disintegrates into liquid filaments which, in turn, are broken, creating droplets. The characteristic break-up time is very short, which means a rapid atomization of the majority of the liquid. The droplets are produced in very small sizes, several orders of magnitude smaller than the diameter of injection.

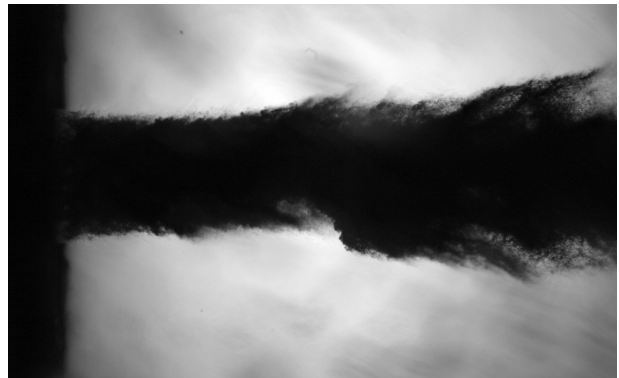
a) Cold flow at instant *t*

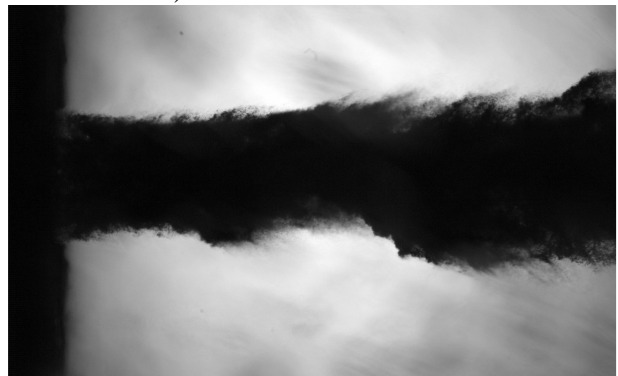
b) Cold flow at instant *t* +0.13ms

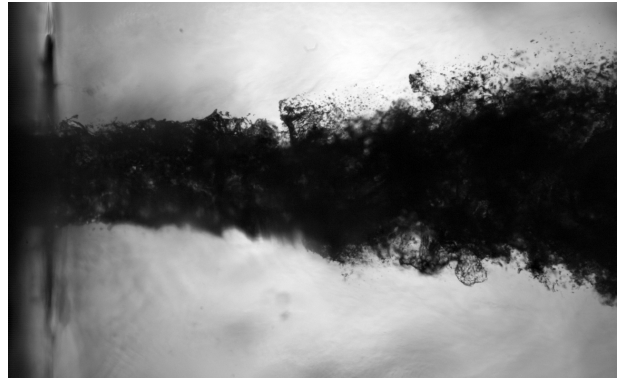
c) Reacting case at instant *t* +0.13ms

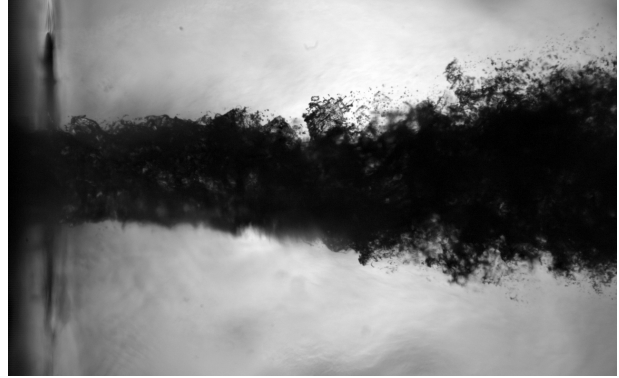
d) Reacting case at instant *t* +0.13ms

Figure 7: Instantaneous image pairs of cold (a-b) and reacting (c-d) flows, recorded at 7.5 kHz with resolution 1280 x 800. Field of view: 5.1*d* x 3.2*d*, image center is at axial distance *x=2d*.

Even if $Re_d$ and $We_G$ were similar, shadowgraphs of the cold and the reacting LOX spray exhibited morphological differences. Indeed the edge of the cold LOX spray was composed of a mist of small droplets ($D<D_{min}$) composing the streaks coming from the LOX core peeling. Those droplets of few pixels are barely noticeable at the scale of figure 7.a and 7.b. Those small droplets, that can be seen on figure 8, with a higher magnification, were surrounding the LOX core and hiding bigger structures. Whereas in the reacting case, this mist of small droplets ($D< D_{min}$) was not present and bigger liquid elements, detaching directly from the LOX core were revealed, like the one in the center of igure 7.c, on the upper edge of the LOX core. The absence of those small droplets is probably due to the flame, which is attached to the lip of the LOX post (Candel et al, 2006,) and evaporates them as soon as the jet exits from the injector. Thus the optical setup used for drop-sizing the reacting flow was not suited to measure the smaller droplet size in the cold flow.

Moreover, the atomization of the liquid fibres could be followed from two successive images of the hot flow whereas it was not possible in the cold case, with the same resolution. This indicates that in the cold flow, liquid structures were moving faster than in the reacting flow. Indeed, the cold spray is constituted of smaller droplets, which are more likely to be accelerated by the outer gas velocity, due to their smaller inertia, than bigger droplets in the reacting case, as mentionned by Chigier and Mc Creath, 1974.

In order to visualize the small liquid structures in the cold flow, we performed high magnification (x5) high speed imaging with a Questar QM100 long distance microscope. At this scale, the recording rate of the camera was increased up to 49 kHz in an attempt to follow the liquid elements in a couple of successive images, as illustrated on figure 8. But for an axial distance *x<10d*, like in figure 7, it was not possible to track the liquid particles because they were moving too fast. We estimated from figure 8 the liquid structure velocity of the cold spray at a distance corresponding to ROI C and D (see bottom of figure 3). Fragments of various sizes could travel on a distance larger than half of the image. We estimated that the velocity of those fragments was about 50 m/s, which was several times faster than in the reacting case. This value was in agreement with the gas phase velocity measured by Gicquel and Vingert, 2000 with a PDPA, at a similar location, in cold conditions too, but with gaseous $N_2$ instead of gaseous He. With the optical setup of the present experiment, the continuous light source induced a pixel shift on shadowgraphs, due to the droplet velocity, thus it created a blurring effect on droplet images and drop-sizing was impossible, even with this high magnification setup.

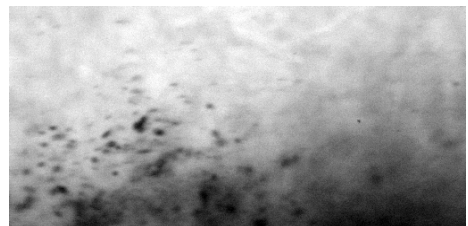

a) Axial distance x=12*d*.

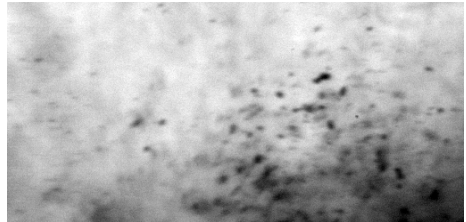

b) Axial distance x=12*d*.

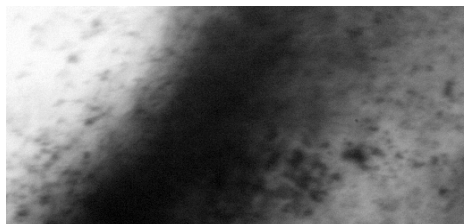

c) Axial distance x=10*d*.

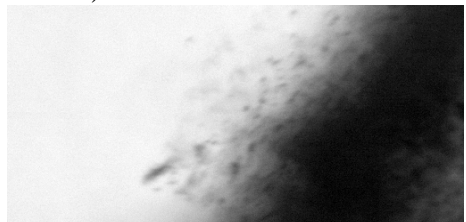

d) Axial distance x=10*d*.

Figure 8: Examples of successive image pairs (a-b and c-d), recorded at 49 kHz obtained with the high magnification setup for the cold flow. Field of view 0.4*d* x 0.2*d*.

3.2 Time averaged images

The time-averaged small-sized pictures have been put together in order to obtain an overall view of the spray on figure 9. Images were recorded with the same optical setup, except that single images of the cold flow were smaller in height to increase the recording rate of the camera. The contrast of images has been enhanced to distinguish the gas and the liquid phase. At first sight, it seems that the cold LOX spray seems shorter than in the reacting case. This observation has been made by several authors who estimated the dense LOX core length of cold flows and reacting jets in similar injection conditions (Davis and Chehroudi, 2007 and Hardi et al., 2014). The dense LOX core can be defined as the axial length of the liquid jet from the injection plane to the breakup of the first oxygen structures. In such a dense spray, the flame front is situated at the periphery of the spray sheath, as indicated by Chigier and Mc Creath, 1974. No burning takes place in the dense LOX core and the main reaction zone remains at the spray periphery. This kind of burning spray can last for long distances from the nozzle, probably more than in the cold flow.

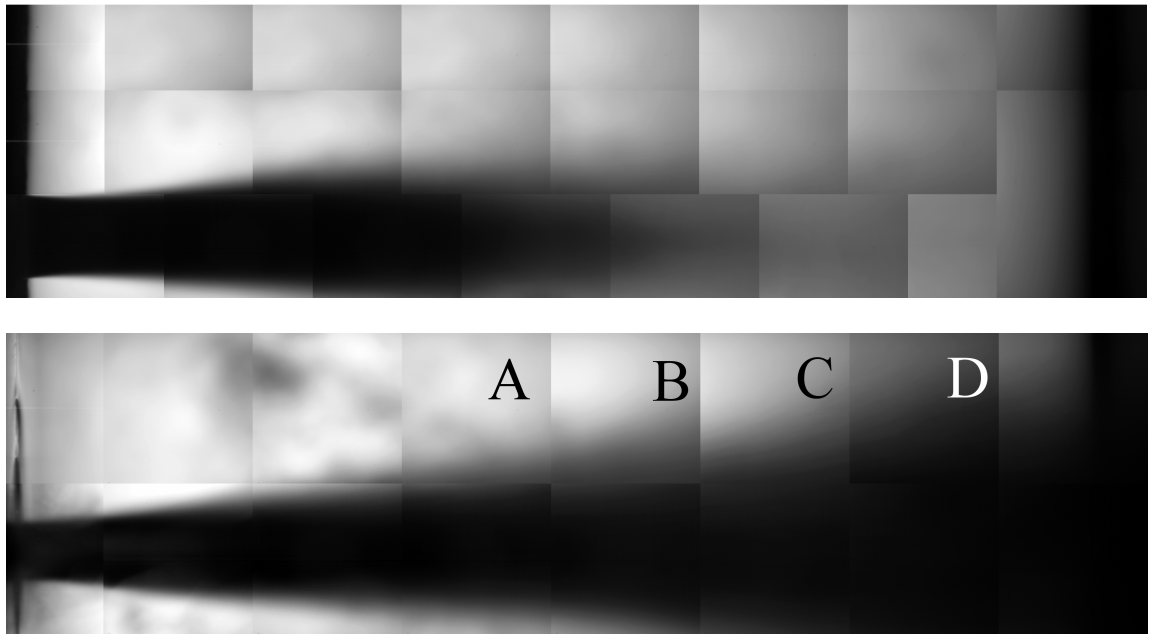

Figure 9: Time averaged images of the spray in cold flow (top) and reacting conditions (bottom). Field of view: 15$d$ x 4$d$.

3.3 Mean diameters and probability density functions $f_n$

Drop size measurements were performed in the reacting case only, from 512 x 512 images, in the ROI A to D, as described in section 2.2 and illustrated on figure 9, corresponding to a radial distance of $r/d=2$ and axial distances $x/d$ from 6 to 12, considering the image center. The Sauter mean diameter $D_{32}$, which is commonly used in combustion studies, and the arithmetical mean diameter $D_{10}$, are presented in figure 10. We superimposed the mean spray image, with the same scale on the horizontal axis, for illustration purpose. For small-size images recorded in ROI A and B, droplets were mainly measured in the bottom half part of images whereas for measurements in ROI C and D, droplets covered almost the entire image field. We only performed imaging processing in ROI A to D because in other small sized images, the particle density was either too high or too low, and thus not exploitable. The droplet concentration $Cv$ could be estimated in the ROI D where the droplet repartition was the most homogeneous. $Cv$ is defined as the ratio of the mean number of droplet per image to the measurement volume of the imaging technique, which can be obtained from the depth-of-field calibration procedure (see figure 5). The droplet concentration was found to be $Cv \approx 0.06$ drop.mm$^{-3}$ in ROI D, for the reacting case. This value is low and probably not representative of the concentration on the jet axis because ROI D is on the spray edge and relatively far from the nozzle where most of small droplets have already been vaporized.

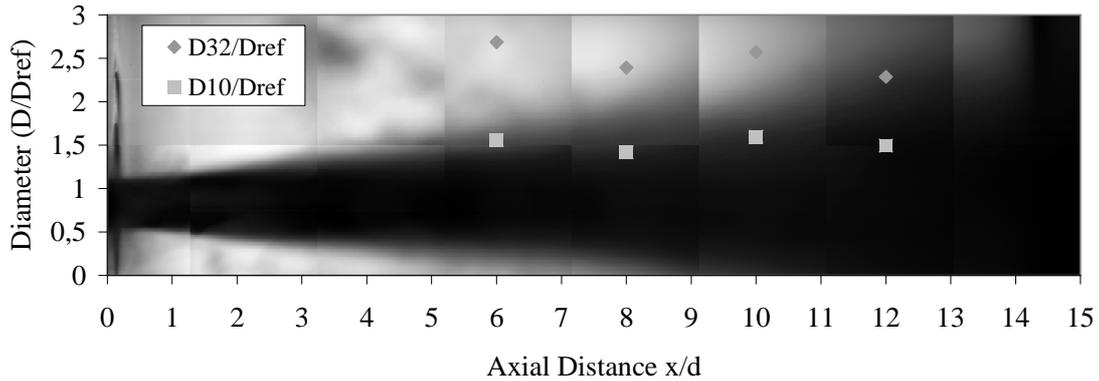

Figure 10: Axial evolution of mean diameters, reacting case.

Figure 10 shows that the mean diameter evolution towards the injection axis appeared to fluctuate in the atomization zone. However the Sauter mean diameter showed a slight decrease towards smaller diameters, as the axial distance was increased. This tendency was confirmed by the axial evolution of the drop-size distribution $f_n$, presented on figure 11. Gicquel and Vingert, 2000, measured the $D_{32}$ in the same conditions with a PDPA, and they found that $D_{32}/D_{ref} \approx 1.55$ at $r/d = 2$ and $x/d = 6$ (in ROI A).

This value is not in agreement from $D_{32}/D_{ref} = 2.7$ here but it is close to the $D_{10}/D_{ref} = 1.55$ value measured in the present study at the same measurement location than Gicquel and Vingert, 2000. This large discrepancy on $D_{32}$ could be explained by the presence of large liquid structures, at this location in the spray, where the droplets are not spherical. Indeed most of them are probably rejected by PDPA inducing a low validation rate for PDPA (30%). However, further from the injector where droplets are more spherical, Gicquel and Vingert, 2000 measured $D_{32}/D_{ref} \approx 2.45$ at $r/d = 2$ and $x/d = 18$, with a higher validation rate. This value is close to the one measured in the present study $D_{32}/D_{ref} \approx 2.3$ at the same radial location but at $x/d = 12$ (ROI D).

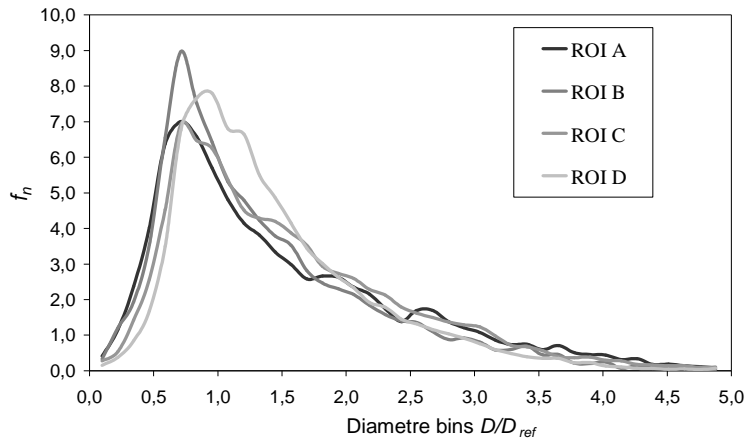

Figure 11: Drop size distribution (*pdf*) evolution towards the injection axis in ROI A to D, reacting case.

The *pdf* exhibited a large shape with a slow decrease towards the biggest diameters. The drop size distribution seemed to be translated towards the biggest diameter as the axial distance from the injector increased. This evolution, often observed in burning sprays, is probably due to the evaporation of the droplets induced by combustion. The smallest droplets ($D/D_{ref}<1.0$) evaporate more rapidly than large droplets ($D/D_{ref}>1.5$), even if the so-called $D^2$ evaporation law is not applicable in such turbulent flames.

3.4 PTV velocity measurements and diameter/velocity correlations

For PTV velocity measurements, the sorting on the depth of field $\chi_{max}$ had to be less restrictive otherwise the droplet number was not large enough (<5000) to obtain reliable statistics. The $\chi_{max}$ value was set to 0.15 mm for PTV, which results in large number of droplets (140076 droplets for ROI A with 577 image pairs). The same image pairs for PTV as for the drop-sizing method were processed. We noticed that this sorting did not affect the Sauter mean diameter more than 5% of the measured value presented in figure 10.

The droplet size/velocity correlations are presented on figure 12 for the ROI A to D to show the axial evolution. Globally, the droplet velocities decreased by a factor of ≈3 from $x/d=6$ to $x/d=12$ and the deceleration was maximum between ROI B and ROI C. The velocity of the smallest droplets ($D/D_{ref}<2$) decreased more rapidly than for the biggest as the axial distance from the injector increased. Indeed, large droplets ($D/D_{ref}>3$) have more inertia and conserved their velocity whereas small droplets are more sensitive to aerodynamic forces present in this turbulent flame. The size/velocity correlation was found particularly flat in ROI A and ROI B, meaning that all droplets have the same velocity, probably because in this area, liquid elements are close to the first atomization zone and their velocity results mainly from the separation with the LOX core, whatever the droplet size is.

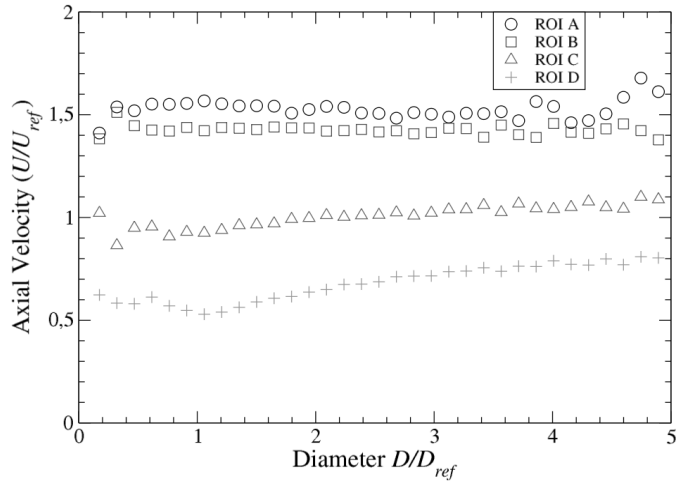

Figure 12: Mean droplet axial velocity measured by PTV as a function of diameter class, reacting case.

3.5 PIV / PTV comparison

In this section we propose to compare the results of two velocity methods: the PTV algorithm previously described and FOLKI-SPIV developed at Onera for PIV applications. The PTV algorithm measures individual droplet movements, which were previously detected by the thresholds of the drop-size algorithm. The FOLKI-PIV method is an iterative gradient-based cross correlation computation which detects movements of patterns in image pairs.

The mean velocity values obtained with PTV and PIV (FOLKI-SPIV) are shown in Table 2, for the reacting case only. Various window sizes (WS) of 33 (not shown here), 63 and 99 pixels wide were used for calculations with FOLKI-SPIV and the larger WS always led to the highest cross correlation score. The mean values were calculated over the whole vector field, with a vector for every pixel for FLOKI-SPIV. For PTV, mean values were calculated over the entire droplet sample, constituted of $N_{d,im}$ droplet number per image (see Table 2), over the same 577 image pairs. We noticed a good agreement between FOLKI-SPIV and PTV mean values, except in ROI A where images showed a large part of gas phase, as can be seen on figure 9. Such a comparison between commercial PTV and PIV algorithms was performed by Goldsworthy et al., 2011 on Diesel spray shadowgraphs. They found a good agreement between PTV and PIV analysis applied directly on shadowgraphs and also with classical PIV images obtained with a laser sheet illumination. Their comparison relied mainly on mean and RMS velocity values obtained with both methods. They found a good agreement in areas where clear patterns of well defined droplets were imaged.

|  | ROI A | ROI B | ROI C | ROI D |
|---|---|---|---|---|
| **PTV** | | | | |
| Mean $U/U_{ref}$ | 1,68 | 1,39 | 0,88 | 0,53 |
| Mean $V/V_{ref}$ | 0,38 | 0,35 | 0,23 | 0,16 |
| $N_{d,im}$ | 74,7 | 95,0 | 136,0 | 106,9 |

| | **FOLKI-SPIV** | | | | | | | |
|---|---|---|---|---|---|---|---|---|
| | Mask | No mask | Mask | No Mask | WS63 | WS99 | WS63 | WS99 |
| Mean $U/U_{ref}$ | 1,64 | 1,29 | 1,33 | 1,31 | 0,72 | 0,74 | 0,51 | 0,52 |
| Mean $V/V_{ref}$ | 0,33 | 0,30 | 0,25 | 0,44 | 0,27 | 0,26 | 0,14 | 0,14 |
| Cross correlation *score* | 0,83 | 0,51 | 0,83 | 0,68 | 0,72 | 0,75 | 0,77 | 0,79 |
| Average valid image pairs | 556 | 509 | 555 | 524 | 539 | 550 | 549 | 553 |

Table 2: Comparison of mean velocities in ROI A to D, measured with PTV and FOLKI-SPIV, with $\Delta t$=0.04 ms, in the reacting case.

We compared radial profiles of the velocity fields obtained by PTV and PIV on the same image pairs, with $\Delta t$=0.04 ms. On figure 13, the mean axial velocity is plotted versus the vertical axis of the image, where $y/d$=0 corresponds to the image bottom. For FOLKI-SPIV, the mean axial velocity was computed every $\Delta y$=0.04$d$, over the whole vector field. For PTV, the mean axial velocity was calculated over all droplets whose gravity center was within a horizontal window of dimensions 2$d$ by 0.04$d$. We can see on figure 13 that results are in a good agreement for ROI C, D and B (not shown). Both methods provided similar velocity profiles when images showed clear patterns of droplet images. For ROI A, there was a good agreement between PIV and PTV only when $y/d$≤0.5, thus where droplets were mainly detected. When $y/d$>0.5, both methods diverged due to a lack of droplets in the upper part of the image. In this zone, PTV seemed to overestimate the axial velocity, due to a lack of droplet samples, whereas FOLKI-SPIV underestimated the displacement, which was likely an effect of the image noise or background that led to a null displacement and thus to a bias toward lower displacement. We noticed that for $y/d$=0.05, the number of particles per image pair $N_{d,im}$, in the horizontal rectangle (of dimensions 2$d$ x 0.04$d$), was about 3.5 drop/image pair for ROI A. This value is the minimum $N_{d,im}$ for which PTV and PIV are in a good agreement with this setup. We noticed that $N_{d,im}$ ≈3.5 drop/image pair in ROI B as well, at $y/d$=3.5 where PIV and PTV started to diverge. Goldsworthy et al., 2011, also noticed that on the edge of a Diesel spray, the mean velocities measured by PTV and PIV showed large discrepancies. They also observed that PTV consistently yielded to higher velocities than for PIV on the spray edges. Thus PIV and PTV velocity measurements in ROI A where $y/d$>0.5 couldn't be considered as reliable. However it was interesting to see that both algorithms agreed well in region of the spray where the dispersed phase is in majority.

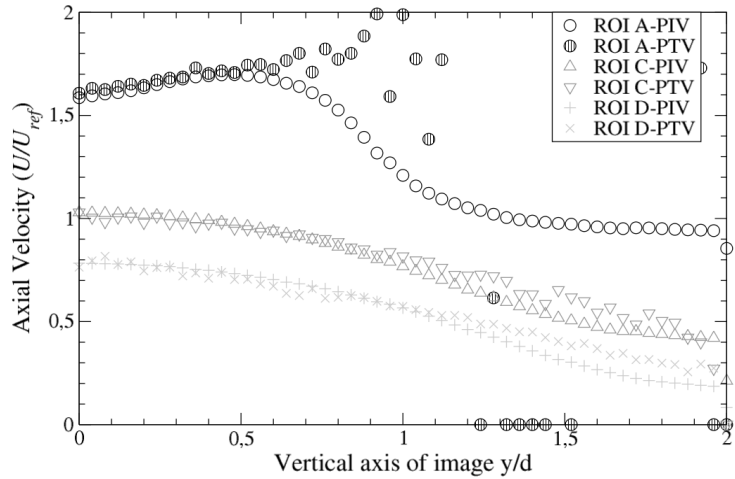

Figure 13: Comparison of radial velocity profiles measured with FOLKI-SPIV and PTV in the reacting case. The vertical axis origin is the image bottom.

We show on figure 14 the overall vector field obtained with FOLKI-SPIV for ROI A to D in the reacting case. Vectors are shown every 17 pixels. For ROI C and D, vectors were obtained in the entire image, because the droplet repartition was quite homogeneous. In ROI A and B, we noticed that the zero centered and normalized pre-processing, applied by FOLKI-SPIV, enhanced the image texture and induced vectors in the gas phase area. Even if this kind of method could be applied to the gas phase, as Johnson et al., 2013 did to characterize the soot mass flux in an atmospheric exhaust plume. In our case it's hard to interpret this velocity measured by FOLKI in the area where gas and liquid were mixing. As the goal of this study is to characterize the dispersed phase velocity only, we chose to apply a mask to eliminate the part of the spray where the gas phase is in majority. The mask was obtained by binarizing the mean image in ROI A and B (see figure 9), at a threshold level ≈0.5 of the maximum grey level, corresponding to the region where PTV and FOLKI-SPIV agreed together. This mask was applied in ROI A and B where the gas phase constituted the majority of pixels and we noticed that in ROI A the cross correlation score increased to 0.83 with the mask (see table 2).

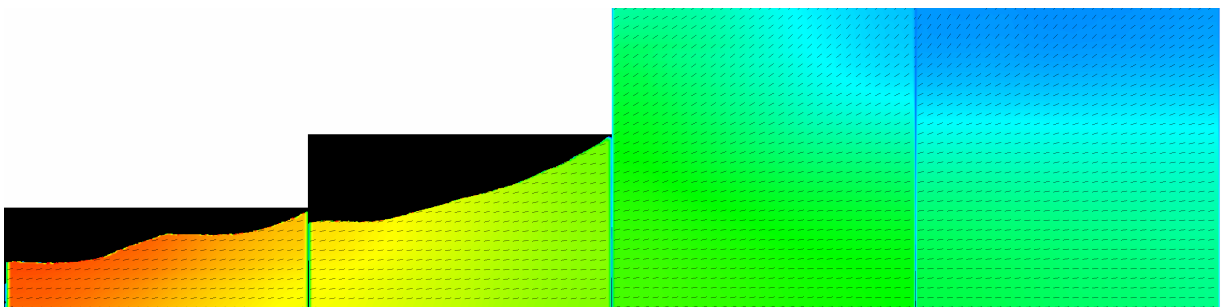

Figure 14: Axial mean velocity field (colorscale) superimposed with the norm of the vector field. ROI A and B computed with a mask, ROI C and D full frame, reacting case.

4 Conclusion and perspectives

In this paper, we showed that high-speed imaging can be a powerful instrument for characterizing cryogenic burning sprays. The optical setup is easier to handle than a PDPA and less sensitive to the shape of liquid elements in such a jet flame. However, PDPA would be better suited to cold conditions because the spray is constituted of smaller and faster droplets. Both methods give the droplet size and velocity correlations which are essential to understand the dynamics of the dispersed phase. Experiments were conducted on the Mascotte test bench, at Onera Palaiseau. This work completes the existing database on the A-10 operating point. A LOX jet was injected with GH2 through a coaxial injector, in subcritical conditions in a 1MPa combustion chamber. Shadowgraphs of the atomization area were performed, where liquid fragments detach from the LOX jet to disintegrate in smaller elements. Imaging was thus more adapted than Phase Doppler techniques in this area where a majority of droplets are not spherical. The morphology of the LOX jet at the injector exit was illustrated with small sized images, recorded at high frequency, for 'similar' cold and reacting conditions, in terms of $Re_d$ and $We_G$ numbers. The cold LOX jet was constituted by an envelope of small droplets ($D<D_{min}$) around the LOX core whereas in the reacting case, those small droplets were not present and bigger liquid structures were revealed. The flame filters the smallest structures by vaporizing them since the LOX post exit. The droplet sizes of the spray were obtained in 4 ROIs of the spray in the first and secondary atomization zone. The Sauter mean diameter evolution with axial distance showed a slight decrease towards the smallest diameters from $x/d=$ 6 to 12, at a radial distance of $r/d=2$. Far from the injector ($x/d=18$), the Sauter mean diameter in ROI D ($x/d=12\pm1$) was found similar to the one measured by Gicquel and Vingert, 2000, with a PDPA. However, close to the injector, at $x/d=6$, where droplets are not spherical, size measurements were not in agreement and difficult to compare because the PDPA validation rate was very low in this area. The probability density functions *pdf* showed some features of such a burning spray: a large shape which seemed to be translated towards large diameters, as axial distance from injector increased, which is probably due to droplet vaporization. Velocity measurements were obtained by a PTV and a PIV based algorithms, both showing that droplet velocities decreased by a factor of 3 from $x/d=6$ to $x/d=12$. Droplet size/velocity correlations were obtained by combining results of the drop-size algorithm and PTV algorithm. The smallest droplets were slowed more rapidly than large droplets, as the axial velocity increased, because they are more sensitive to the aerodynamic forces in the turbulent flame. In ROI A, the closest to the injector exit, all droplets, whatever their size, were ejected with the same velocity. This particular area of the spray could indicate the first atomization zone where the droplet velocity comes from the liquid jet. Velocity of the dispersed phase was also obtained with a PIV algorithm applied directly to shadow images. PIV was performed with FOLKI-SPIV, developed at Onera. This method can be useful to obtain information on such a LOX spray, which is not compatible with most of seeding particles. PIV and PTV algorithms agreed well in terms of $U$, $V$ mean velocity and radial profiles in ROI B, C and D. FOLKI-SPIV gave better results in terms of cross correlation score and agreement when using a large interrogation window of 99 pixels wide. Where the droplet density was not high enough, both methods diverged, and we could not conclude in areas where the gas phase is in majority. Thus we chose to mask those areas where the algorithm was less reliable PIV in terms of score.

The drop size and velocity of such a dense and polydisperse spray, in reacting or cold conditions cannot be studied with a single optical diagnostic. Bringing together data obtained with different methods usually gives complementary information. Imaging diagnostics are more suited to characterize non spherical liquid elements in the atomization zone, particularly in reacting conditions because the flame filters the smallest droplets. But the higher droplet density and velocity in the cold spray prevent from size or velocity measurements with the same imaging setup. Phase Doppler techniques would be more suited to the cold spray where droplets are moving faster. Images of the spray in cold and hot conditions showed that droplet size and velocity were different and the perspectives of this work are to study the dynamics of the atomization process in the cold case, by means of optical diagnostics: a Phase Doppler interferometer would be used to measure the drop size. Results would be compared to the hot case and further numerical simulations. High speed imaging techniques could be also used to characterize the spray dynamics in the first atomization zone to identify spatial and temporal modes with decomposition methods, such as proper orthogonal decomposition, to capture the jet oscillations which produce large liquid elements.

Nomenclature

| | | |
|---|---|---|
| *a* | Actual radius of a droplet | mm |
| $C_0$ | Contrast of the droplet image | Dimensionless |
| $Cv$ | Droplet concentration | Drops/mm$^3$ |
| *d* | Injector diameter | mm |
| *D* | Droplet diameter | mm |
| $D_{min}$ | Minimum measurable droplet diameter | mm |
| $D_{ref}$ | Reference diameter | mm |
| *fn* | Numerical probability density function | Dimensionless |
| *J* | Momentum flux ratio | Dimensionless |
| *Lb* | Penetration length | mm |
| $N_{d,im}$ | Number of droplet per image | Dimensionless |
| *Pc* | Pressure in the combustion chamber | MPa |
| *Pinj* | Injection Pressure | MPa |
| $p_c(O2)$ | Critical pressure of oxygen | MPa |
| $Re_d$ | Reynolds number based on LOX diameter | Dimensionless |
| *ROF* | Propellant mixture ratio | Dimensionless |
| $r_{meas}$ | Measured droplet radius | mm |
| *Tinj* | Injection Temperature | K |
| $T_c(O2)$ | Critical temperature of oxygen | K |
| *t0* | Time origin | s |
| *t* | Time | s |
| *U* | Axial velocity | m/s |
| $U_{ref}$ | Reference axial velocity | m/s |
| *V* | Vertical velocity | m/s |
| $V_{ref}$ | Reference vertical velocity | m/s |
| $We_G$ | Gaseous Weber number | Dimensionless |
| $\chi$ | PSF half width | mm |
| $\Delta d_m$ | Mean distance between two particles in a single image. | mm |
| $\Delta d_t$ | Mean travel distance of a particle between an image pair | mm |
| $\Delta t$ | Time interval between two images | s |
| $\Delta z$ | Depth-of-focus of imaging system | mm |
| $\sigma$ | Standard deviation of velocity | m/s |